\documentclass[conference]{IEEEtran}
\IEEEoverridecommandlockouts

\usepackage{cite}
\usepackage{amsmath,amssymb,amsfonts}
\usepackage{algorithmic}
\usepackage{graphicx}
\usepackage{textcomp}
\usepackage{xcolor}
\usepackage{booktabs}
\usepackage{pbalance}
\usepackage{tabularx}
\usepackage{booktabs}
\usepackage{makecell}
\usepackage{array}
\usepackage{tikz}
\usepackage{siunitx}
\usepackage{url}

\usepackage{pifont}
\newcommand{\cmark}{\ding{51}} 

\newcolumntype{L}[1]{>{\raggedright\arraybackslash}p{#1}}
\newcolumntype{Y}{>{\raggedright\arraybackslash}X}

\def\BibTeX{{\rm B\kern-.05em{\sc i\kern-.025em b}\kern-.08em
    T\kern-.1667em\lower.7ex\hbox{E}\kern-.125emX}}

\renewcommand{\figurename}{Figure}
\newcommand*\circled[1]{\tikz[baseline=(char.base)]{
            \node[shape=circle,draw,inner sep=1pt] (char) {#1};}}

\newcommand{\projectname}{\textbf{\texttt{AEGIS}}}

\newcommand{\dlp}{DLP}
\newcommand{\ebpf}{eBPF}
\newcommand{\slurm}{Slurm}

\begin{document}

\title{Attestation is All You Need: Toward a Zero-Trust Architecture for HPC AI Agents}

\author{
\IEEEauthorblockN{
Jie Li\IEEEauthorrefmark{1},
Chenxu Niu\IEEEauthorrefmark{2},
Tongyang Wang\IEEEauthorrefmark{1},
Yong Chen\IEEEauthorrefmark{1}
}
\IEEEauthorblockA{\IEEEauthorrefmark{1}
Department of Computer Science, Texas Tech University, Lubbock, TX, USA\\
\{jie.li, tongyang.wang, yong.chen\}@ttu.edu
}
\IEEEauthorblockA{\IEEEauthorrefmark{2}
NVIDIA, Santa Clara, CA, USA\\
chenxun@nvidia.com
}
}

\maketitle

\begin{abstract}
AI agents are beginning to automate data analysis, orchestration, and simulation steering in HPC environments, but they introduce a security problem that existing cluster defenses do not address: the hijacked authorized agent. An adversary can subvert an agent through prompt injection, poisoned shared data, compromised tools, or multi-agent coordination, then exfiltrate sensitive data through legitimate encrypted LLM API traffic. We propose behavioral attestation as a runtime defense for this setting. Our system, \projectname, compiles task-derived behavioral constraints, monitors agent actions with \ebpf-based syscall interception, verifies evidence against constraint profiles, correlates behavior across agents, and triggers \slurm{}-level containment. On the real attestation path, \projectname~detects all four implemented attack classes with sub-second median latency at a 1-second interval and low measured verifier overhead. Controlled experiments show full attack-suite coverage, stronger detection than representative baselines, and no observed false positives on the evaluated benign workflows.
\end{abstract}


\section{Introduction}

The convergence of large language models with high-performance computing is transforming scientific workflows. Autonomous agents are beginning to appear alongside established workflow automation systems for data-intensive science, orchestration, and large-scale experiment management~\cite{deelman2015pegasus,babuji2019parsl,jain2015fireworks,xi2025rise}. These AI agents promise to accelerate discovery, but they also introduce a threat category that HPC security was never designed to address.

The core problem is not that AI agents are untrustworthy. The problem is that they are too trustworthy, specifically, they trust the wrong inputs. Recent work has shown that LLM-integrated applications and tool-using agents are vulnerable to prompt injection and instruction-channel confusion~\cite{greshake2023not,liu2024formalizing,liu2023prompt}. An agent authorized to process genomics data for Project X will faithfully execute whatever instructions it receives, including instructions embedded within that data by an attacker. A prompt injection payload hidden in a \textit{FITS (Flexible Image Transport System)} file header, a malicious command left in a shared \textit{/tmp} directory by a co-located job, or a compromised tool that returns hidden instructions during normal operation represent concrete attack vectors. These are not theoretical concerns; they are the natural consequence of deploying instruction-following systems into environments where untrusted data is routine.

This paper identifies and formalizes a threat that, to our best knowledge, has not been studied in prior work for HPC specifically: \textit{the hijacked authorized agent}. Unlike rogue agents, which authentication blocks, or adversarial agents, which authorization constrains, a hijacked agent operates with the full credentials and privileges of a legitimate user. It does not need to escalate privileges because it already has them. It does not need to bypass access controls because they already permit everything the user can do. From the perspective of every existing monitoring system that validates identity, the agent is legitimate. Yet it exfiltrates data through precisely the channel no HPC system inspects: the encrypted HTTPS connection to its LLM backend.

The attack surface is uniquely HPC. Shared parallel filesystems such as Lustre and IBM Spectrum Scale create injection vectors with no direct web analogue~\cite{lustre2025intro,ibm2024spectrumscale}. Multi-tenant compute nodes scheduled by \slurm{} create co-location injection opportunities through shared temporary storage and kernel resources~\cite{yoo2003slurm}. Tool-use protocols and agent integration layers make it easier to connect agents to external services and local executables, but they also widen the trusted computing base~\cite{mcp2024spec,mialon2023augmented}. Coordinated multi-agent workflows enable distributed exfiltration that evades per-agent reasoning entirely.

Existing defenses are fundamentally ill-equipped against this threat. Authentication succeeds because the agent has valid Kerberos tickets. Authorization succeeds because the agent has legitimate POSIX permissions. Network monitoring is blinded because LLM API traffic is encrypted. Data loss prevention fails because sensitive data can be embedded within whitelisted HTTPS requests. User behavior analytics report nothing because the agent's workflow patterns remain superficially consistent with its authorized role. The defender's toolkit checks identity, but not intent.

We propose \textbf{behavioral attestation} as the foundation for securing AI agents in HPC. Our key insight is straightforward: we cannot prevent all injection attacks, because the attack surface is too large and diverse. However, we can contain their effects by detecting and responding to the behavioral violations that hijacked agents inevitably produce. When an agent accesses files outside its authorized project, connects to non-whitelisted endpoints, invokes unauthorized tools, or exceeds its data budget, attestation detects the violation and enforces containment before larger-scale exfiltration can continue unchecked.

Behavioral attestation differs from prior system-integrity attestation approaches along three fundamental dimensions. First, it provides constraint-based verification rather than probabilistic alerts. Existing agent defenses often rely on classifiers or pattern-matching systems with inherent false positive and false negative rates. Behavioral attestation asks whether an action satisfies the declared policy boundary. Second, it is constraint-based rather than signature-based. Existing defenses often block known-bad patterns, which attackers can evade through obfuscation. Constraints define what is allowed, not what is malicious. Third, it operates at runtime rather than post hoc. Existing monitoring systems often alert after the attack succeeds; behavioral attestation verifies constraints continuously during execution and triggers containment while the session is still active.

This paper makes the following contributions.

\begin{itemize}
    \item We formalize the hijacked authorized agent threat in HPC, identifying \textit{four} distinct attack vectors that exploit unique characteristics of HPC infrastructure.
    \item We introduce behavioral attestation as a constraint-based runtime security primitive that extends zero-trust principles from identity to agent behavior.
    \item We design and implement \projectname, a complete zero-trust architecture encompassing constraint specification, \ebpf-based attestation, policy verification, cluster-wide correlation for coordinated attack detection, and automated containment through \slurm{} integration.
    \item We demonstrate that \projectname~detects all four implemented attack vectors on the real attestation path, achieves full coverage in the controlled attack suite, outperforms representative baselines, and maintains low measured runtime overhead on HPC hardware.
\end{itemize}

\section{Background}

This section provides background on zero-trust security, AI agents and workflow automation, the security posture of contemporary HPC systems, and prompt injection in LLM-integrated applications.

\subsection{Zero-Trust Architecture}

Zero-trust architecture replaces coarse network perimeters with continuous policy decisions over identities, devices, services, and resources. The original zero-trust framing emphasized the idea that location on a trusted network should not be treated as a security boundary~\cite{kindervag2016no}. NIST SP 800-207 later formalized zero trust as an architecture centered on explicit authentication, authorization, and continuous evaluation~\cite{rose2020zero}. NIST SP 800-207A extended this framing to cloud-native applications, highlighting workload identity, application telemetry, and fine-grained service-to-service enforcement~\cite{chandramouli2023zero}.

These ideas are increasingly relevant to HPC, but they do not directly solve the autonomous-agent problem. Existing zero-trust deployments focus on who is requesting access and under what context. In an AI-agent setting, that is necessary but insufficient: a legitimate, correctly authenticated agent may still behave outside its authorized task after being hijacked by adversarial inputs.

\subsection{AI Agents and Scientific Workflow Automation}

Classical agent research studies autonomous entities that perceive, reason, and act to achieve goals~\cite{wooldridge2009introduction}. Modern LLM-based agents extend this model with tool use, planning, reflection, and long-horizon execution~\cite{schick2023toolformer,yao2022react,wang2023voyager,mialon2023augmented,xi2025rise,masterman2024landscape}. These systems increasingly rely on external tools and protocol-mediated service access, which expands both their capability and their attack surface~\cite{mcp2024spec}.

HPC environments already depend on large-scale automation through workflow and scheduling systems such as Pegasus, Parsl, FireWorks, and \slurm{}~\cite{deelman2015pegasus,babuji2019parsl,jain2015fireworks,yoo2003slurm}. AI agents are a natural next step in that automation stack because they can adapt plans at runtime rather than executing only pre-declared Directed Acyclic Graphs (DAGs). However, this same flexibility means that an attacker can steer them through data, tool outputs, or communication channels that the surrounding infrastructure still treats as legitimate workflow inputs.

\subsection{Security in HPC}

Contemporary HPC sites remain strongly identity- and scheduler-centric. Authentication often relies on Kerberos, SSH, and federated research identity infrastructure such as CILogon~\cite{neuman1994kerberos,basney2013cilogon}. Authorization is enforced through scheduler policy, POSIX permissions, ACLs, and shared storage semantics. Storage and execution environments are optimized for performance and collaboration rather than for strict tenant isolation; in practice, sites commonly rely on \slurm{}, Lustre, and IBM Spectrum Scale or related shared filesystems~\cite{yoo2003slurm,lustre2025intro,ibm2024spectrumscale}.

This model works reasonably well for human-driven batch computing, but it leaves an important gap for autonomous agents. Once a user launches an agent under valid credentials, the agent inherits the user's authority across files, tools, and network endpoints. The system verifies identity and resource ownership, but it does not continuously verify whether the resulting sequence of actions remains faithful to the user's task.

\subsection{Prompt Injection and Agent Security}

Prompt injection has emerged as a fundamental security problem for LLM-integrated applications because language models do not inherently distinguish trusted instructions from adversarial content embedded in data~\cite{greshake2023not,liu2023prompt,liu2024formalizing}. The broader risk picture for LLM systems includes misuse, prompt leakage, jailbreaks, and downstream tool abuse~\cite{perez2022red,weidinger2022taxonomy,owasp2025llm}. Universal adversarial suffixes further show that LLM guardrails are brittle even when developers try to enforce instruction hierarchies purely in the prompt~\cite{zou2023universal}.

A growing literature explores partial defenses and evaluation frameworks. Spotlighting improves source separation between trusted instructions and untrusted data~\cite{hines2024defending}, while StruQ proposes structured prompting to separate instructions from data payloads~\cite{chen2025struq}. ToolEmu and AgentDojo provide controlled environments for surfacing safety failures and adversarial behavior in tool-using agents~\cite{ruan2023identifying,debenedetti2024agentdojo}. These works establish the threat and improve evaluation practice, but they do not directly address the cluster-level runtime enforcement problem studied in this paper.

\section{Threat Model}

\subsection{The Hijacked Agent Threat}

\begin{figure}[!t]
    \centering
    \includegraphics[width=\columnwidth]{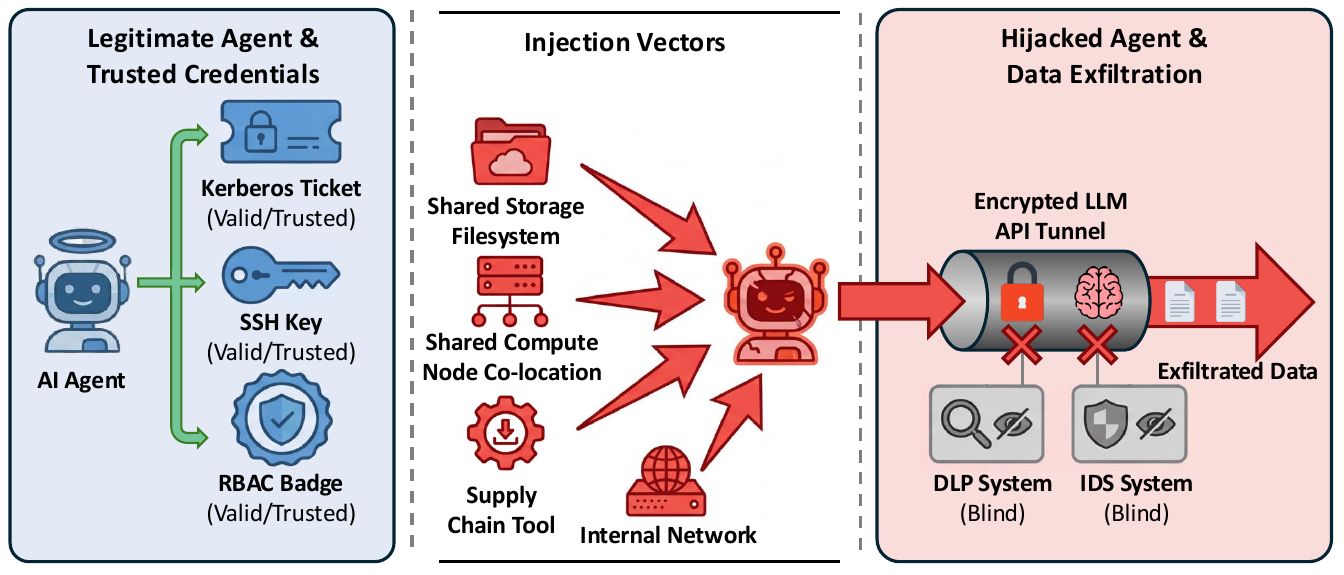}
    \caption{The hijacked authorized agent threat. An agent with valid credentials is subverted through injection attacks. The hijacked agent exfiltrates data through encrypted LLM API channels that are difficult for traditional Data Loss Prevention (DLP) and Intrusion Detection System (IDS) to interpret.}
    \label{fig:threats}
\end{figure}

We identify the most dangerous threat to HPC environments deploying AI agents: \textit{the hijacked authorized agent}, as illustrated in Figure~\ref{fig:threats}. Unlike malicious agents, which are blocked by existing authentication and authorization mechanisms, a hijacked agent operates under the full credentials and privileges of a legitimate user. It is not an intruder; it is a trusted principle that has been turned.

This threat arises from known prompt-injection and tool-mediated control-transfer vulnerabilities in LLM-integrated applications~\cite{greshake2023not,liu2023prompt,liu2024formalizing,perez2022red}. Benchmarks such as ToolEmu and AgentDojo further show that tool-using agents remain brittle under adversarial inputs and environmental manipulation~\cite{ruan2023identifying,debenedetti2024agentdojo}. An attacker crafts inputs through data files, tool outputs, shared documents, or collaborative channels that subvert the agent's instruction-following behavior. The agent, now under adversarial control, executes commands that are externally attributable to a legitimate workflow.

A hijacked agent possesses \textit{four} properties that make it uniquely dangerous in HPC. \circled{1} \textbf{Full credential inheritance}: the agent operates with the user's Kerberos tickets, SSH keys, and scheduler permissions. No privilege escalation is needed; the agent already has access. From the access control system's perspective, every action is authorized. \circled{2} \textbf{Cross-project filesystem access}: HPC shared filesystems are typically organized by user, not by project. A user authorized on Projects A, B, and C grants their agent simultaneous access to all three datasets. A hijacked agent can traverse project boundaries that would require separate authorization in a properly segmented system. \circled{3} \textbf{Appearance of legitimacy}: the agent runs under an authorized user identity, invokes authorized tools, and follows authorized workflow patterns. Traditional anomaly detection fails because the agent's observable behavior including job submission, file I/O, and network access is consistent with its authorized role. \circled{4} \textbf{Exfiltration through the LLM API channel}: the agent communicates with its LLM backend via HTTPS API calls. Sensitive data can be encoded into prompts or tool outputs and transmitted through this channel, which is encrypted, whitelisted, high-bandwidth, and attributable to normal operation.

This exfiltration vector is difficult for traditional \dlp{} systems to handle. The problem is not just that the traffic is encrypted, but that the sensitive action itself often looks operationally plausible unless the defender reasons about the relationship between the agent's task and its observed behavior.







\subsection{Unique Properties of HPC Agent Injection Attacks}

\begin{figure}[!t]
    \centering
    \includegraphics[width=\columnwidth]{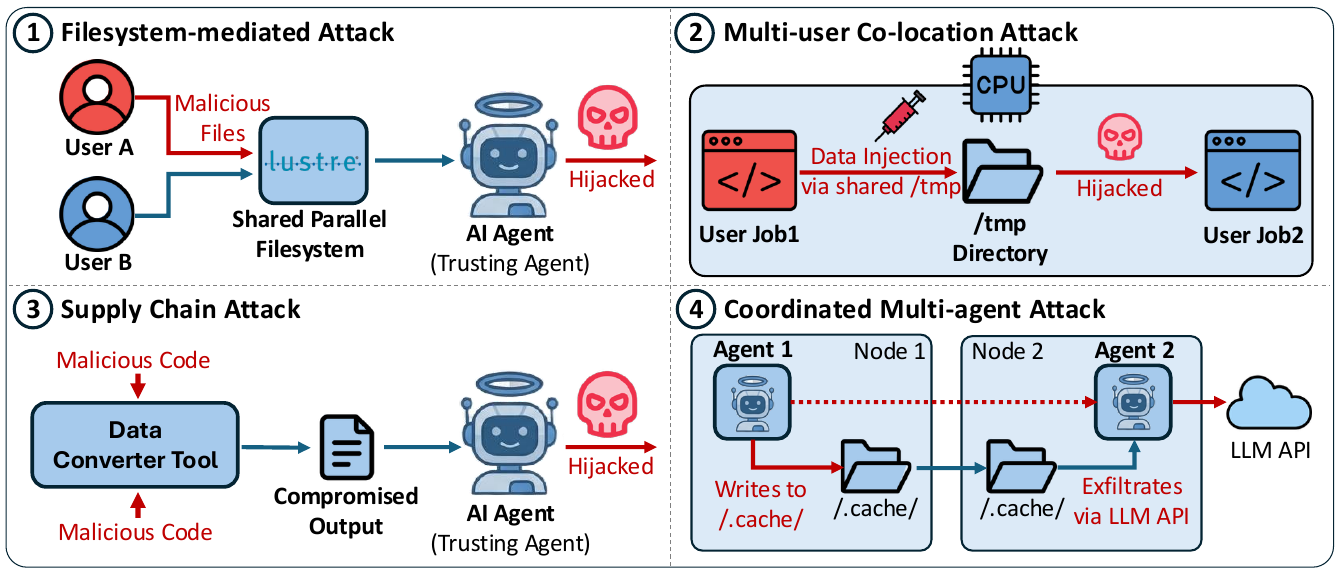}
    \caption{Four HPC-specific injection attack surfaces. (1) Filesystem-mediated: poisoned data in shared parallel filesystems. (2) Co-location: shared \textit{/tmp} on multi-tenant compute nodes. (3) Supply chain: compromised tools or services in the agent control loop. (4) Coordinated: multi-agent covert exfiltration over shared resources.}
    \label{fig:injections}
\end{figure}

Agent injection attacks in HPC exploit \textit{shared infrastructure} in ways that differ from the web-centric scenarios emphasized in most prior prompt-injection literature. These properties arise from the structure of the environment itself: shared filesystems, scheduler-driven co-location, rich tool connections, and multi-agent execution, as illustrated in Figure~\ref{fig:injections}.

\circled{1} \textbf{Filesystem-mediated injection} arises from a core property of HPC systems: shared storage. Filesystems such as Lustre introduce injection surfaces that have no direct counterpart in web environments. An attacker with access to a shared project space can plant adversarial content (e.g., poisoned metadata, hidden instructions embedded in log outputs) that a target agent later reads as if it were legitimate scientific data. Because HPC workflows depend on treating shared data as trustworthy, this assumption cannot be easily relaxed without disrupting normal operation.

\circled{2} \textbf{Multi-user co-location injection} stems from how schedulers pack jobs onto shared nodes. Jobs from different users often run side by side, enabling application-level side channels. An attacker can write adversarial content to shared locations such as \textit{/tmp}, which a co-located agent may subsequently read. This does not require access to the target’s project data, only that both jobs land on the same node.

\begin{figure*}[!t]
  \centering
  \includegraphics[width=0.95\linewidth]{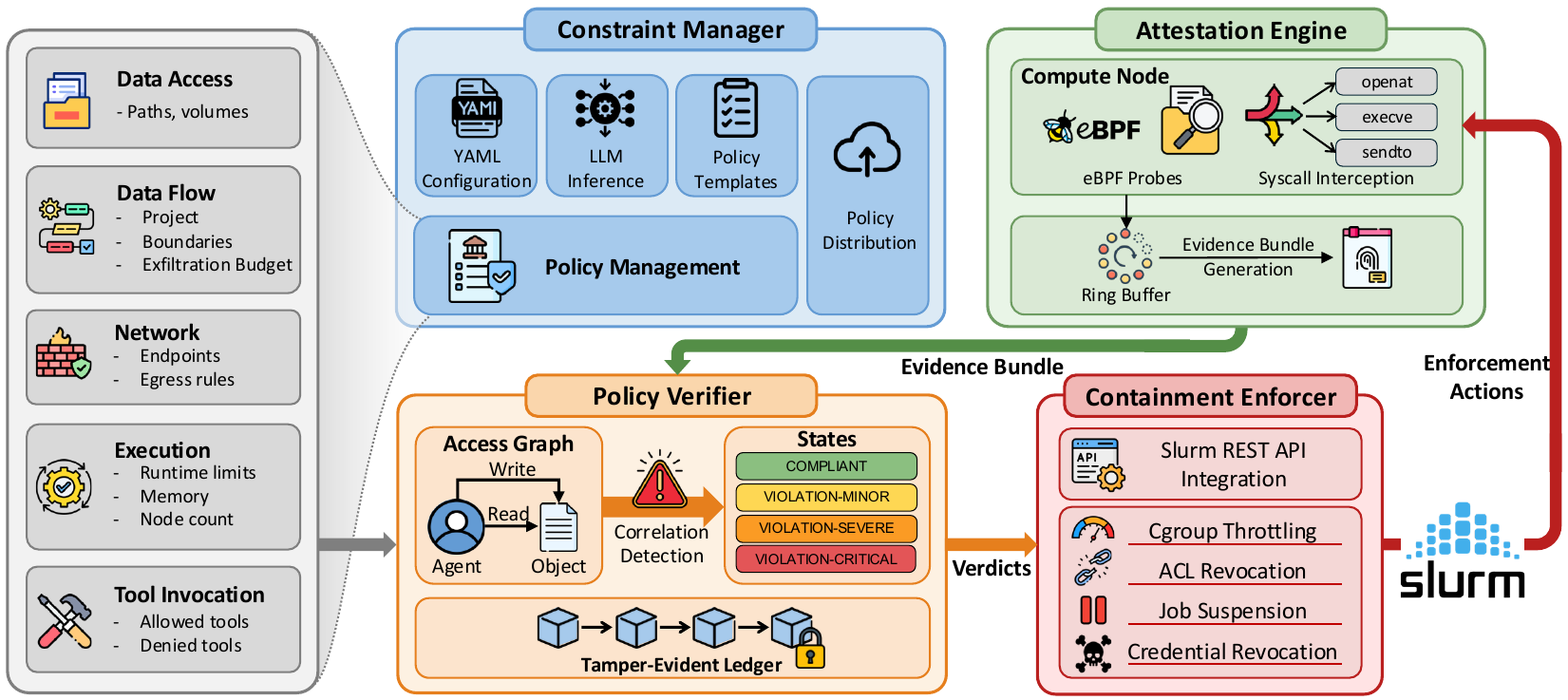}
  \caption{Overview of the \projectname~system architecture. Four core components: Constraint Manager, Attestation Engine (per-node, \ebpf-based), Policy Verifier (centralized), and Containment Enforcer (via Slurm REST API).}
  \label{fig:aegis_arch}
\end{figure*}

\circled{3} \textbf{Supply chain injection via agent skills} reflects risks introduced by tool ecosystems such as OpenClaw, LangChain, and MCP servers. These components resemble traditional software supply chains: if a tool is compromised, it can inject malicious instructions directly into the agent’s control loop via its outputs. Since the agent intentionally invoked the tool, the resulting behavior appears indistinguishable from normal execution.

\circled{4} \textbf{Coordinated multi-agent exfiltration} captures attacks that span multiple agents. Compromised agents across different users or projects can cooperate to move data covertly. One agent may stage sensitive data in a shared location, while another retrieves it and exfiltrates it through its own LLM API channel. Each agent’s behavior, viewed in isolation, appears benign; the attack only becomes apparent when considering interactions across agents.

\subsection{Adversarial Capabilities}

We consider an adversary who can craft inputs that manipulate an agent's instruction-following behavior through prompt injection or tool-mediated poisoning. The adversary may have access to shared filesystems, collaborative channels, or co-located jobs, and may observe network metadata such as traffic timing and volume, though not the encrypted contents of LLM API traffic. The adversary does not begin with root access to the cluster and does not directly compromise the scheduler, verifier, or LLM provider infrastructure.

We also assume that the defender controls the cluster control plane, including the scheduler integration, verifier, and node-level collector deployment. The adversary may hijack agents and persist across sessions, but does not compromise the verifier's signing keys or the cluster's administrative trust anchors.

\subsection{Why Existing Defenses Fail}

Authentication and authorization both succeed because the agent is operating with valid credentials and legitimate permissions. However, this is precisely what allows the attack to remain difficult to distinguish from normal use. Network monitoring is ineffective because LLM API traffic is encrypted and explicitly whitelisted. \dlp{} mechanisms are also poorly matched because sensitive data can be embedded within otherwise legitimate prompts or tool outputs. User-behavior analytics struggle because the agent's actions remain broadly consistent with its assigned role. Even sandboxing provides limited protection, since the agent must retain access to the filesystem, tools, and network in order to perform its intended task.

Taken together, these limitations motivate \textbf{behavioral attestation}: continuous verification that the agent's actions conform to its authorized behavioral boundary, not merely that its identity is valid.

\section{Behavioral Attestation for AI Agents}
\label{aegis}

We propose behavioral attestation as the foundation for securing AI agents in HPC. We implement this concept in \projectname~(\textbf{A}ttestation-based \textbf{E}nvironment for \textbf{G}uarding \textbf{I}njection-vulnerable \textbf{S}ystems).

\subsection{Overview}

Figure~\ref{fig:aegis_arch} illustrates the \textit{four} core components of \projectname~that operate across the HPC cluster. The \textbf{Constraint Manager} parses and compiles behavioral constraint profiles into an internal policy representation. It supports explicit specification, task inference, and policy templates. Profiles are signed and bound to the agent's Slurm job ID. The \textbf{Attestation Engine} runs as a daemon on each compute node, intercepting agent system calls via \ebpf~probes. It produces signed attestation evidence bundles at configurable intervals, transmitted to the verifier over mutually authenticated gRPC. The \textbf{Policy Verifier} is a centralized service that evaluates evidence against constraint profiles, producing verdicts ranging from \texttt{COMPLIANT} to \texttt{VIOLATION-CRITICAL}. It issues random challenges to prevent delayed reporting and maintains a shared access graph across all attested agents on the cluster, enabling cross-agent correlation to detect coordinated attacks such as covert channels via shared filesystem paths. The verifier correlates write-read patterns across agent sessions: if Agent A writes to a path and Agent B reads from the same path within a sliding time window, this triggers a covert channel alert. All decisions are logged to a tamper-evident audit ledger. The \textbf{Containment Enforcer} translates verdicts into enforcement actions via the Slurm REST API. It supports cgroup throttling for minor violations, ACL revocation for moderate violations, job suspension for severe violations, and termination with credential revocation for critical violations.

These components implement behavioral attestation, a fundamentally new concept that differs from prior approaches along \textit{three} axes. Prior approaches provide detection through probabilistic mechanisms such as ML-based classifiers with inherent false positive and negative rates. Behavioral attestation provides deterministic constraint verification with optional signature augmentation. Prior approaches use signature-based policies that block known-bad patterns which can be evaded through obfuscation. Behavioral attestation uses constraint-based policies that define what is allowed, where violations are unambiguous. Prior approaches operate post-hoc, alerting after the attack succeeds when damage is already done. Behavioral attestation operates at runtime, detecting and containing violations during execution.

The key insight underlying our approach is that we cannot prevent all injection attacks because the attack surface is too large and too diverse. However, we can \textbf{detect and contain the effects of hijacked agents by attesting to behavioral constraints}. \projectname~uses a hybrid detection model: constraint-based verification serves as the primary mechanism and ensures that agent actions conform to declared behavioral boundaries. An optional signature layer augments detection with known injection patterns for specific attack vectors. The constraint layer alone achieves 75\% detection as shown in the ablation study, and signatures boost this to 100\% for attacks that produce recognizable patterns.

\subsection{Behavioral Constraint Profiles}

Each agent receives a behavioral constraint profile at deployment, declarations of legitimate behavior derived from the agent's authorized task, not signatures of malicious behavior. Constraints span five dimensions: \textit{data access} (allowed/denied paths, read-only paths, volume limits), \textit{network} (whitelisted endpoints, egress budgets), \textit{tool invocation} (allowed/denied tools), \textit{execution} (runtime limits, memory limits, node restrictions), and \textit{data flow} (project boundaries, cross-agent isolation). Constraints are evasion-resistant: an attacker cannot make an unauthorized action authorized through obfuscation; the constraint either permits the action or it does not. Cross-agent constraints specify whether the agent may share data with other agents, access co-located agent outputs, or participate in inter-agent communication channels.

We support three modes of constraint derivation. Explicit specification allows users to declare constraints directly through a declarative YAML-based language. Task inference uses an LLM to analyze the agent's task description and automatically generate appropriate constraints. Policy templates provide pre-built constraint sets for common HPC agent patterns such as data analysis, simulation steering, and ML training.

The constraint profile is cryptographically bound to the agent's execution context. We link the profile to the agent's Slurm job ID, creating a verifiable binding between the authorized behavior and the running process. This prevents an attacker from substituting a more permissive profile.

\subsection{Attestation Evidence}

The Attestation Engine monitors agent behavior through \ebpf-based syscall interception~\cite{gregg2019bpf}. The engine hooks six syscall tracepoints: \texttt{sys\_enter\_openat} for file access detection, \texttt{sys\_enter\_read} and \texttt{sys\_enter\_write} for byte counters, \texttt{sys\_enter\_connect} for network connections, \texttt{sys\_enter\_sendto} for network egress tracking, and \texttt{sys\_enter\_execve} for tool invocations. Events are emitted via the BPF ring buffer to userspace, and Section~\ref{results} quantifies both worst-case syscall overhead and end-to-end verifier cost on the deployed path. The probe tracks per-PID state including cumulative bytes read and written, and connection counts.

At configurable intervals (default: 1 second), the Attestation Engine produces a signed attestation evidence bundle containing the agent's access log, data volume counters, hash of the agent's process state, and a cryptographic signature. The evidence bundle is transmitted to the Policy Verifier for evaluation.

\begin{figure*}[t]
  \centering
  \includegraphics[width=\textwidth]{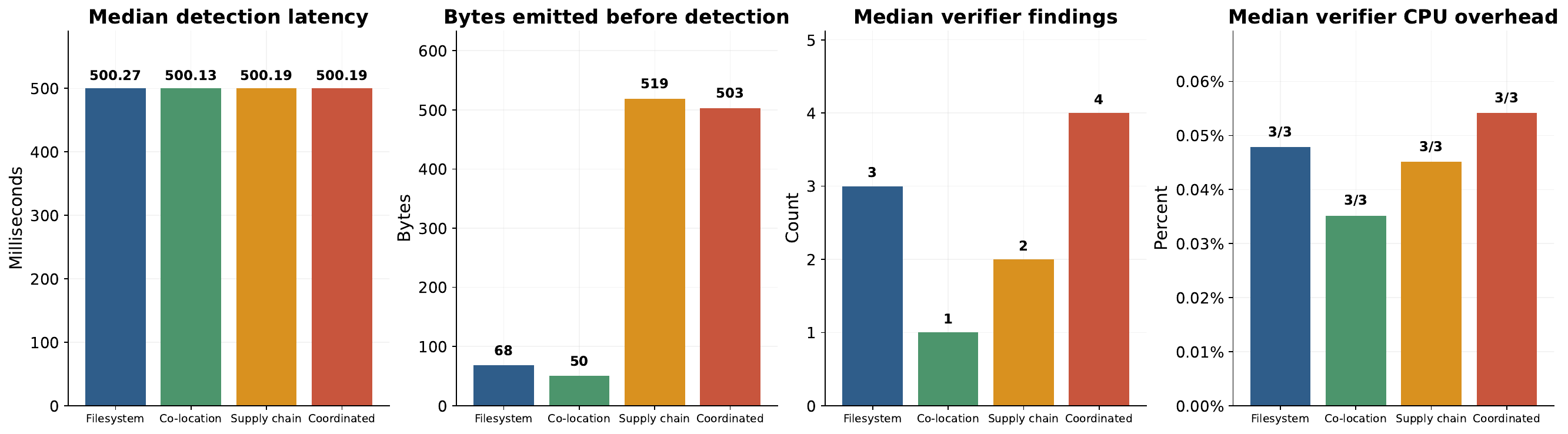}
  \caption{Real-path outcomes at a 1\,s attestation interval. \projectname~
  detects all four attacks in 3/3 trials with median detection latency of
  500.13--500.27\,ms, 1--4 findings in the first detecting attestation
  bundle, and verifier CPU overhead below 0.055\%.}
  \vspace{-10pt}
  \label{fig:attack-results}
\end{figure*}

\subsection{Policy Verification}

The Policy Verifier evaluates evidence bundles against constraint profiles, producing verdicts that indicate the severity of any violations. The verdict scale ranges from \texttt{COMPLIANT} (no violations) through \texttt{VIOLATION\_MINOR}, \texttt{VIOLATION\_MODERATE}, \texttt{VIOLATION\_SEVERE}, to \texttt{VIOLATION\_CRITICAL} (severe threat requiring immediate action).

The verifier maintains a cluster-wide access graph that tracks all agent filesystem and network interactions across all compute nodes. This enables cross-agent correlation to detect coordinated attacks. The verifier correlates write-read patterns across agent sessions: if Agent A writes to a path and Agent B reads from the same path within a sliding time window, this triggers a covert channel alert. The time window and correlation thresholds are configurable. All verification decisions are logged to a tamper-evident audit ledger. The ledger uses a cryptographic hash chain, enabling post-hoc forensic analysis and detection of log tampering.

\subsection{Containment Enforcement}

The Containment Enforcer translates verification verdicts into enforcement actions through the Slurm REST API. Four escalation levels are supported. For minor violations, the enforcer applies cgroup-based rate limiting, throttling the agent's CPU and memory resources. For moderate violations, the enforcer revokes filesystem ACLs, isolating the agent from shared storage. For severe violations, the enforcer suspends the Slurm job, pausing agent execution. For critical violations, the enforcer terminates the job and revokes credentials, ensuring the compromised session cannot be reused.

\subsection{Optional Signature Layer}

For deployments requiring maximum detection coverage, \projectname~includes an optional signature-based augmentation layer. This layer detects known prompt injection patterns in tool outputs and inter-agent messages using pattern matching against a curated database of injection signatures such as hidden instruction patterns and base64-encoded commands. The signature layer is disabled by default and must be explicitly enabled; it is not required for constraint-based detection to function. In the controlled synthetic ablation, the non-signature mechanisms detect 3 of 4 attacks (75\%), and enabling injection signatures raises detection to 4 of 4 by catching the crafted supply-chain attack.

\subsection{Implementation Details}

\projectname~is fully implemented as a modular system. The implementation consists of approximately 4,300 lines of code. The C \ebpf{} probe provides kernel-side syscall hooks, and the Python collector reads from the ring buffer to assemble evidence bundles. The centralized verifier maintains the cluster-wide access graph and tamper-evident audit ledger, while the evaluation harness instantiates the baseline DLP, auditing, analytics, and sandbox comparators used in Section~\ref{results}. The \slurm{} integration layer translates verifier decisions into containment actions through the REST API. The implementation is released with the artifact for reproducibility.

\section{Experimental Evaluation}
\label{results}

\subsection{Methodology}

We evaluate the deployed stack on an AMD EPYC 9754-based node (256 cores at
2.25\,GHz, 512\,MB shared L3, and 1536\,GB system memory) running Rocky Linux
9.6 with Linux kernel 5.14.0. This platform is representative of current HPC
compute nodes. The \ebpf-based attestation engine is compiled with
\texttt{clang} and performs kernel-side syscall interception, while the
verifier executes the full \projectname~policy pipeline including
behavioral-constraint checking, cross-agent correlation, and containment
decision generation.

Our evaluation separates \emph{real-path measurements} from
\emph{controlled security studies}. The real-path measurements exercise the
implemented probe, collector, and verifier stack on the EPYC platform to
measure direct probe overhead, end-to-end detection latency, and per-attack
behavior. Unless stated otherwise, real-path results use a 1-second attestation
interval and report medians over three repetitions per attack.

The controlled studies use the same policy semantics and attack families, but
execute them in a deterministic harness that exposes larger attack, ablation,
and workload matrices than those that are convenient to stage on a single node. We use
this harness for complete attack coverage, specialized-component ablation, the
benign-workflow false-positive study, the scaling analysis, and the baseline
comparison. Controlled attack outcomes are deterministic for a given trace; for
baseline timing we analyze each trace ten times and report mean analysis time.
The benign suite covers six workflows and 42 benign actions, including
authorized multi-agent collaboration, budget-edge LLM reporting, and tool-heavy
post-processing.

\begin{figure*}[t]
  \centering
  \includegraphics[width=\textwidth]{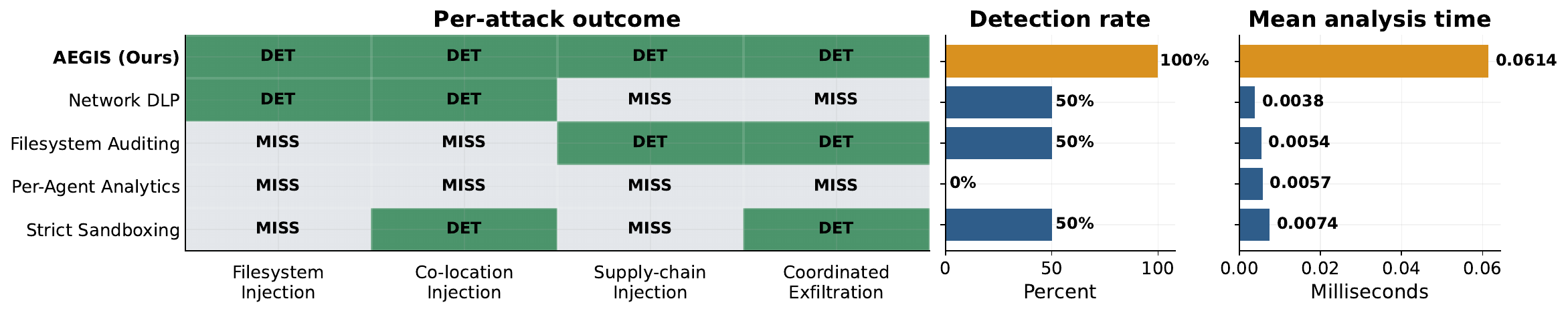}
  \caption{Baseline comparison on identical attack traces. \projectname~is the
  only defense that detects all four attacks; the simpler baselines detect at
  most two of four. Mean analysis time remains sub-0.1\,ms in all cases.}
  \vspace{-10pt}
  \label{fig:baseline-comparison}
\end{figure*}

\subsection{Real-Path Detection Results}

Figure~\ref{fig:attack-results} summarizes the real-path detection results.
\projectname~detects all four implemented attack classes in 3/3 trials and in
the first post-injection attestation cycle. At the default 1\,s attestation
interval, the median detection latency is 500.27\,ms for filesystem-mediated
injection, 500.13\,ms for co-location injection, 500.19\,ms for supply-chain
injection, and 500.19\,ms for coordinated exfiltration. These latencies are
approximately half of the attestation interval because the attacks are injected
midway through a bundle window. Median data emitted before detection is 68\,B,
50\,B, 519\,B, and 503\,B, respectively. The first detecting attestation
bundle contains a median of 3, 1, 2, and 4 verifier findings, respectively,
which reflects how many policy dimensions each attack violates. Median verifier
CPU overhead stays below 0.055\% in all cases.

The narrow latency spread across the four real-path attacks indicates
stable verifier behavior, while the emitted-byte and finding-count differences
reflect attack semantics rather than detector instability. The real-path results also show that \projectname~detects attacks through the
intended policy dimensions rather than through a single catch-all rule.
Filesystem injection triggers denied-path and unauthorized-endpoint violations,
co-location injection is detected through network policy, supply-chain
injection combines sensitive-file access with exfiltration policy, and
coordinated exfiltration triggers both denied-path violations and the
cross-agent covert-channel rule.

\subsection{Controlled Security Coverage}

To complement the real-path measurements, we run the full four-attack suite in
the controlled harness. Table~\ref{tab:controlled-coverage} summarizes these
results. In this setting, all four attacks succeed in hijacking the target
agent and exfiltrating data, and \projectname~detects all four, yielding a 100\%
detection rate over 4/4 completed attacks. The controlled suite produces 16
total findings and 1158\,B of exfiltrated payload across the four attacks. The
simpler filesystem and co-location attacks each trigger one detection, whereas
the supply-chain and coordinated attacks each trigger seven detections because
they violate multiple policy dimensions simultaneously. The coordinated attack
also raises an explicit covert-channel finding, confirming that the
cluster-wide access graph is necessary for this class of multi-agent attack.

\begin{table}[t]
  \caption{Controlled security coverage across the four attack classes.}
  \label{tab:controlled-coverage}
  \centering
  \footnotesize
  \setlength{\tabcolsep}{4pt}
  \begin{tabular}{L{0.5\columnwidth}cccc}
    \toprule
    Attack & Det. & Find. & \makecell{Time\\(ms)} & \makecell{Exf.\\(B)} \\
    \midrule
    Filesystem-mediated injection        & \cmark & 1 & 0.258 & 68  \\
    Multi-user co-location injection     & \cmark & 1 & 0.038 & 50  \\
    Supply-chain injection via skills    & \cmark & 7 & 0.089 & 519 \\
    Coordinated multi-agent exfiltration & \cmark & 7 & 0.104 & 521 \\
    \bottomrule
  \end{tabular}
\end{table}

These controlled results matter for two reasons. First, they provide a fully
reproducible security-coverage view that is not tied to one hardware
configuration. Second, they show that the real-path detections are not
accidental: the same attack classes remain detectable when exercised in a
deterministic harness with the same policy semantics.

\subsection{Baseline Comparison}

\begin{table}[t]
  \caption{Controlled ablation of specialized detection mechanisms. Each row removes one verifier component and reports whether the matching attack class remains detectable.}
  \label{tab:ablation-summary}
  \centering
  \scriptsize
  \setlength{\tabcolsep}{2.6pt}
  \begin{tabularx}{\columnwidth}{@{}L{0.38\columnwidth}ccccc@{}}
    \toprule
    Configuration & \makecell{Volume\\limit} & \makecell{Sensitive\\file} & \makecell{Covert\\channel} & \makecell{Injection\\signature} & \makecell{Detection\\rate} \\
    \midrule
    Full \projectname & $\checkmark$ & $\checkmark$ & $\checkmark$ & $\checkmark$ & \textbf{100\%} \\
    No volume limits & $\times$ & $\checkmark$ & $\checkmark$ & $\checkmark$ & 75\% \\
    No sensitive detection & $\checkmark$ & $\times$ & $\checkmark$ & $\checkmark$ & 75\% \\
    No covert-channel correlation & $\checkmark$ & $\checkmark$ & $\times$ & $\checkmark$ & 75\% \\
    No injection signatures & $\checkmark$ & $\checkmark$ & $\checkmark$ & $\times$ & 75\% \\
    Minimal path-only policy & $\times$ & $\times$ & $\times$ & $\times$ & 0\% \\
    \bottomrule
  \end{tabularx}
\end{table}

Figure~\ref{fig:baseline-comparison} compares \projectname~against four
representative alternatives on the same deterministic attack traces. Network
DLP, filesystem auditing, and strict sandboxing each detect two of the four
attacks (50\%), while per-agent analytics misses all four. \projectname~is the
only defense that detects every attack class. The richer analysis does incur
higher processing cost than the simpler baselines, but the absolute cost
remains small: \projectname~averages 0.0614\,ms of analysis time per trace,
compared with 0.0038--0.0074\,ms for the baselines.

These results highlight why \projectname~needs all four policy dimensions. DLP
sees egress but cannot reason about encrypted traffic sent to whitelisted
services. Filesystem auditing sees sensitive paths but misses network-only
violations. Strict sandboxing blocks only the attacks that cross a hard
isolation boundary, and per-agent analytics misses coordinated behavior
entirely because the individual agents remain locally plausible. In contrast,
\projectname~combines constraint checking, signature augmentation, and
cross-agent correlation in a single verifier.

\begin{figure*}[t]
  \centering
  \includegraphics[width=\textwidth]{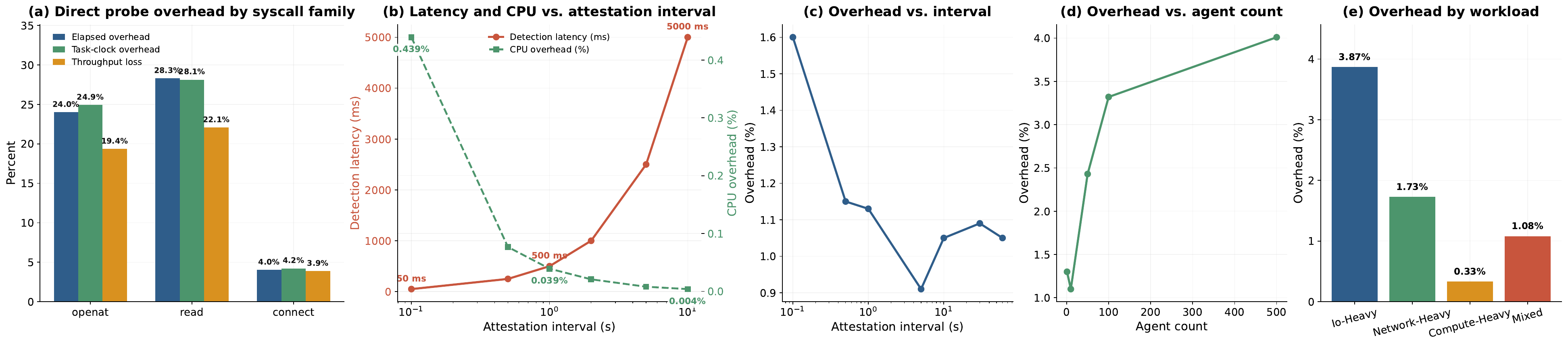}
  \caption{Performance summary. From left to right: direct probe overhead by syscall family, deployed detection latency and verifier CPU overhead versus attestation interval, controlled overhead versus attestation interval, controlled overhead versus agent count, and controlled overhead by workload type. The deployed path stays below 0.44\% CPU and falls to 0.039\% at the default 1\,s interval, while 14 of 16 controlled configurations remain below 5\% overhead.}
  \label{fig:performance-summary}
\end{figure*}

\subsection{Ablation}

Table~\ref{tab:ablation-summary} reports the single ablation result that we
use in the paper, because the controlled harness isolates the contribution of
each mechanism more cleanly than the overlapping deployed-stack checks. In this
setting, full \projectname~detects all four controlled attacks. Disabling
volume limits, sensitive-file detection, covert-channel correlation, or
injection signatures drops detection from 100\% to 75\% by allowing exactly
the matching attack class to pass, while a minimal path-only policy detects
none of the attacks. The controlled ablation therefore supports the intended
interpretation: declarative constraints provide the main security boundary, and
specialized checks add orthogonal coverage where generic path constraints are
insufficient.

This breakdown is also consistent with the real-path behavior discussed above.
On the deployed stack, network policy is the mechanism that catches co-location
injection, and cross-agent correlation is what preserves detection for
coordinated exfiltration even when the local profile is otherwise permissive.
The ablation evidence therefore shows both overlap and specificity: several
checks reinforce one another, but distinct attack classes still depend on
specific policy dimensions.

\begin{table*}[t]
  \caption{Benign-workflow suite used to assess false positives. Across six
  workflows, 42 benign actions and 70 logged events produce zero alerts. The
  table also summarizes which benign stress dimensions each workflow exercises.}
  \label{tab:false-positive-summary}
  \centering
  \small
  \setlength{\tabcolsep}{6pt}
  \begin{tabularx}{\textwidth}{@{}L{4.6cm}cccYc@{}}
    \toprule
    Workflow & Agents & Actions & Logged Events & Benign Stress Exercised & Alerts \\
    \midrule
    Genomics Data Analysis             & 1 & 6 & 11 & Shared-path data access; longer trace                           & 0 \\
    ML Training Pipeline               & 1 & 7 & 12 & Tool chaining; longer trace                                     & 0 \\
    Multi-Agent Collaboration          & 2 & 8 & 14 & Shared paths; multi-agent coordination; longer trace            & 0 \\
    Simulation Steering                & 1 & 6 & 9  & Longer trace                                                    & 0 \\
    Budget-Edge LLM Reporting          & 1 & 8 & 13 & Authorized near-budget egress; longer trace          & 0 \\
    Tool-Heavy Climate Postprocessing  & 1 & 7 & 11 & Tool chaining; longer trace                                     & 0 \\
    \bottomrule
  \end{tabularx}
\end{table*}

\subsection{Performance}
Figure~\ref{fig:performance-summary} presents five complementary
performance views. Subplot~(a) reports direct probe overhead by
syscall family, subplot~(b) reports the deployed latency and verifier CPU
tradeoff versus attestation interval, and subplots~(c)--(e) report the
controlled scaling study across attestation interval, agent count, and
workload mix.

For the direct microbenchmarks in subplot~(a), file-centric syscalls are the
most expensive: elapsed-time overhead is 24.0\% for \texttt{openat} and
28.3\% for \texttt{read}, while network connection monitoring adds 4.04\%
for \texttt{connect}. Because each run stresses a single syscall family
continuously, these numbers should be interpreted as pessimistic upper bounds
on probe cost rather than application-level slowdown. The same runs show
throughput losses of 19.4\% for \texttt{openat}, 22.1\% for \texttt{read},
and 3.9\% for \texttt{connect}.

Subplot~(b) shows that end-to-end verifier overhead is much lower on the full
attestation path. All four attacks remain detectable across the tested interval
range from 100\,ms to 10\,s. Detection latency follows the attestation
interval, increasing from 50.2\,ms at 100\,ms to 500.2\,ms at 1\,s and
5.00\,s at 10\,s, while verifier CPU overhead falls from 0.439\% to 0.039\%
and then to 0.0039\%. This makes the deployment tradeoff explicit: the default
1\,s setting sits near a practical knee of the curve, delivering sub-second
detection at about 0.04\% measured verifier CPU overhead, while moving to
500\,ms roughly halves latency at nearly twice the CPU cost and relaxing to
2\,s halves CPU again but doubles latency.

Subplots~(c)--(e) show the controlled scaling trends that are inconvenient
to measure directly on the single-node EPYC setup. In that study, 14 of 16
tested configurations remain below 5\% overhead. In the controlled interval
sweep at 10 agents and mixed workload, overhead stays between 0.91\% and
1.60\% across the full 100\,ms to 60\,s range, with the default 1\,s point
at 1.13\%. In the separate agent-count sweep at a fixed 1\,s interval, the
mixed workload incurs 1.10\% overhead for 10 agents, 2.43\% for 50 agents,
3.32\% for 100 agents, and 4.01\% for 500 agents. In the workload-type sweep
at 10 agents and 1\,s, overhead ranges from 0.33\% for compute-heavy jobs to
3.87\% for I/O-heavy jobs. We interpret these results as controlled scaling
trends rather than deployment measurements.

\subsection{Benign Workloads}

Table~\ref{tab:false-positive-summary} summarizes the benign-workflow study.
The expanded benign-workflow suite produces no observed false positives.
Across six workflows, 42 benign actions, and 70 logged events,
\projectname~reports zero alerts. The suite is not limited to trivially clean
traces: it includes standard single-agent HPC pipelines, authorized
multi-agent collaboration, budget-edge LLM reporting, and tool-heavy
post-processing, covering both workflow breadth and benign edge cases close to
the intended policy boundary.

The result shows that no false
positives were observed on the evaluated benign suite, not that false
positives are impossible. Nevertheless, the combination of zero observed false
positives and the broadened workflow coverage suggests that declarative
behavioral constraints can remain precise enough for practical HPC agent
workloads.

\section{Related Work}

\subsection{Zero trust and identity-centric access control.}
Zero trust architectures replace coarse network perimeters with explicit
authentication, authorization, and continuous policy evaluation over users,
services, and resources. NIST SP 800-207 established this model at the system
architecture level~\cite{rose2020zero}, and NIST SP 800-207A extended it to
cloud-native deployments with stronger emphasis on workload identity,
application-level telemetry, and fine-grained enforcement in distributed
services~\cite{duckworth2023spiffe}. \projectname~follows the same continuous-trust
intuition, but shifts the enforcement target from user or service identity to
agent behavior. In our setting, a valid identity is not sufficient: the
relevant question is whether a running agent remains within the behavioral
constraints implied by its authorized HPC task.

\subsection{LLM-agent security and prompt injection.}
Recent work has shown that LLM-integrated applications and tool-using agents
are vulnerable to indirect prompt injection, tool abuse, and other forms of
instruction-channel confusion. Greshake et al. demonstrated that untrusted
content can steer real LLM-integrated applications by exploiting the mismatch
between data and instructions~\cite{greshake2023not}. Liu et al.
formalized prompt-injection attacks and evaluated attack and defense strategies
across models and tasks~\cite{liu2024formalizing}. ToolEmu and AgentDojo then
provided structured environments for stress-testing tool-using agents and
surfacing safety failures under adversarial conditions~\cite{ruan2023identifying, debenedetti2024agentdojo}.
These efforts are important for characterizing the threat model, but they are
primarily benchmark- and application-oriented. \projectname~addresses a different
problem: how to constrain and verify the behavior of deployed agents at runtime
on a shared cluster, including filesystem, network, and cross-agent effects
that emerge only in multi-tenant execution.

\subsection{Remote attestation and runtime integrity.}
Remote attestation has traditionally focused on establishing trust in platform
state. Sailer et al. linked attestation to access-control policy
enforcement~\cite{sailer2004attestation}; Haldar et al. proposed semantic
remote attestation for higher-level properties beyond simple code
measurement~\cite{haldar2004semantic}; and PRIMA extended integrity
measurement toward policy-relevant information-flow properties~\cite{jaeger2006prima}.
Kil et al. argued that practical attestation must reason about dynamic runtime
properties rather than only boot-time state~\cite{kil2009remote}, while more
recent work such as ARI targets mission-execution integrity in cyber-physical
and real-time environments~\cite{wang2023ari}. \projectname~is complementary to this
line of work but changes the attested object. Rather than attesting only code
identity, platform state, or timing properties, \projectname~attests whether an AI
agent's observed actions satisfy behavioral constraints derived from its
authorized job, and it extends that reasoning to cluster-wide correlations
across co-located agents.

\subsection{Data leakage prevention.}
Data leakage prevention systems have been widely studied for data at rest, in
use, and in transit. Shabtai et al. surveyed leakage-detection and prevention
techniques across enterprise channels~\cite{shabtai2012survey}, and Alneyadi et
al. categorized modern DLP systems by their content- and context-analysis
strategies~\cite{alneyadi2016survey}. These systems are effective when leakage can
be recognized from content signatures, protocol context, or policy violations
at a single endpoint or network boundary. However, they are not designed for
the hijacked-agent setting studied here, where the exfiltration channel may be
a legitimate encrypted API and the key violation is behavioral rather than
content-based. In \projectname, the crucial question is not whether a payload matches
a known sensitive pattern, but whether the agent should have performed the
underlying read, write, or network action at all. \projectname~therefore complements
traditional DLP with behavioral attestation and runtime containment rather than
content inspection alone.

\section{Discussion}

\subsection{Deployment Implications}

Our evaluation suggests that behavioral attestation is practical for the class
of HPC agent workloads targeted by \projectname, but its value depends on deployment
discipline rather than on instrumentation alone. The most important lesson is
that agent identity is not a sufficient security primitive for autonomous
scientific workflows. Once an agent is granted legitimate filesystem, network,
and tool access, the relevant question is whether its \emph{observed behavior}
continues to match its authorized task. \projectname~therefore fits naturally into HPC
sites that already operate with strong identity and scheduler control, but need
a runtime mechanism for detecting abuse of legitimate authority.

The real-path results in Section~\ref{results} show that this model can be realized with an
\ebpf-based probe, a verifier-driven policy pipeline, and scheduler-mediated
containment without imposing large steady-state cost. The 1\,s configuration
provides sub-second median detection latency with low measured verifier CPU
overhead, which makes it a practical default for interactive and batch-oriented
agent workflows. More importantly, the experiments show that cluster context is
necessary: the coordinated attack is only visible when the verifier correlates
behavior across agents rather than reasoning about each session in isolation.

In practice, this means \projectname~should be deployed as part of the control plane,
not as an application library. The constraint manager, verifier, and containment
logic need access to job metadata, scheduler actions, and cluster-wide event
streams. The design is therefore most useful in settings where the site can
control both the job launch path and the enforcement path, such as Slurm-based
research clusters, AI-enabled scientific gateways, or managed institutional
agent platforms.

\subsection{Limitations}

The primary limitation of behavioral attestation is that its guarantee is only
as strong as the behavioral constraint profile. If a profile omits an important
resource dimension or encodes an overly permissive boundary, then actions in
that gap are not violations. This is not a flaw in the verifier; it is a policy
specification problem. Templates and inference can reduce operator burden, but
high-assurance deployments still require human review of the resulting
constraints.

A second limitation is response granularity. \projectname~does not prevent the initial
attack step; it detects and contains the resulting policy violation. The size of
that response window is bounded by the attestation interval and the containment
path. Our evaluation shows that the trade-off is favorable at a 1\,s interval,
but the window is not zero, and highly latency-sensitive deployments may need
shorter intervals or stronger inline enforcement for selected resources.

A third limitation is the trust model. The current design assumes a trustworthy
kernel, a trustworthy node-level collector, and an uncompromised scheduler
control plane. \ebpf-based monitoring raises the bar relative to user-space
instrumentation, but it does not solve the problem of a malicious kernel or a
fully privileged attacker on the node. Similarly, the centralized verifier used
in our prototype simplifies cluster-wide correlation, but it is not yet the
final scaling architecture for very large systems with thousands of nodes and
high event rates.

Finally, our evaluation intentionally separates real-path measurements from larger controlled studies. This separation improves methodological clarity but means that some claims in Section~\ref{results} reflect deployment measurements, while others derive from controlled security experiments. This distinction also delineates the boundary of what the current prototype has been physically exercised to support.

\subsection{Future Directions}

Several extensions would strengthen the system. First, constraint generation can
be improved with richer policy templates, stronger provenance for inferred
profiles, and review workflows that make it easier for users and site operators
to validate what an agent is actually allowed to do. Second, the verifier can be
extended with more scalable data structures and distributed aggregation for
cluster-wide correlation. Third, containment can be made more adaptive by tying
verdicts to job QoS controls, storage isolation, and site-specific incident
response policies.

More broadly, behavioral attestation appears useful beyond the specific threat
model studied here. The same core idea, declare what an autonomous actor is
allowed to do, continuously verify observed behavior, and contain violations at
runtime, applies to other multi-tenant AI deployments, including cloud-hosted
agents, scientific gateways, and tool-using assistants operating over shared
enterprise data. HPC is a compelling first target because it combines valuable
data, strong scheduler control, and increasingly autonomous agents, but the
underlying security primitive is more general.

\section{Conclusion}

This paper argues that the main security problem introduced by AI agents in HPC
is not unauthorized access by an outsider, but misuse of legitimate authority by
a hijacked authorized agent. Once an agent is trusted with project data, shared
storage, tools, and network access, conventional authentication and access
control no longer answer the critical question: whether the agent's runtime
behavior still matches its authorized task.

We introduced behavioral attestation as a practical security primitive for this
setting and presented \projectname, a system that compiles behavioral constraints,
monitors agent actions with \ebpf-based syscall interception, verifies evidence
against constraint profiles, correlates behavior across agents, and translates
verdicts into scheduler-level containment actions. The central design choice is
to treat agent security as a runtime verification problem over declared
behavioral boundaries rather than as a purely identity-based or
signature-matching problem.

Our evaluation shows that this approach is effective and practical. On the real
attestation path, \projectname~detects all four implemented attack classes with
sub-second median latency at the default 1\,s interval and low measured verifier
overhead. In the controlled studies, \projectname~achieves full coverage over the
four-attack suite, outperforms representative baselines, preserves zero
observed false positives on the evaluated benign workflows, and shows clear
ablation evidence that the full system derives its coverage from multiple
complementary mechanisms rather than from a single rule.

The broader implication is that securing AI agents requires a shift from
verifying who the agent is to verifying what the agent is doing. For shared HPC
infrastructure, that shift is especially important because the attack surface is
large, the data is valuable, and legitimate workflows already require broad
access to files, tools, and external services. Behavioral attestation does not
eliminate prompt injection or tool compromise, but it does provide a concrete
way to detect and contain their effects before they become cluster-scale data
exfiltration events. We believe this is a necessary systems foundation for the
next generation of AI-enabled scientific computing.

\section{Acknowledgement}
We acknowledge the use of ChatGPT (version GPT-5.3) to assist in revising the manuscript, including improvements to grammar, clarity, and presentation. All technical content, analyses, and conclusions remain the responsibility of the authors.

\balance
\bibliography{references}
\bibliographystyle{IEEEtran}

\end{document}